\documentclass[aps,prl,twocolumn,superscriptaddress,floatfix]{revtex4}
\usepackage{graphicx}
\usepackage{hyperref}
\usepackage{amsmath}
\usepackage{amsfonts}
\usepackage{mathtools,bm}
\usepackage{color}
\usepackage{physics}
\usepackage{amsthm}
\usepackage{pzccal}
\usepackage{multirow}

\theoremstyle{plain}
\newtheorem{thm}{Theorem}[section]

\newtheorem{prop}[thm]{Proposition}

\newcommand{\expo}[1]{e^{#1}} 

\newcommand{\ud}[1]{{#1^{\dagger}}}

\newcommand{\g}[1]{g^{(#1)}}

\begin{document}
\flushbottom
\title{Structure of the Harmonic Oscillator in the space of $n$-particle Glauber correlators}

\author{E. Zubizarreta Casalengua}
\affiliation{Departamento de F\'isica Te\'orica de la Materia Condensada, Universidad Aut\'onoma de Madrid, E-28049, Spain.}

\author{J.~C.~L\'opez Carre\~no}
\affiliation{Departamento de F\'isica Te\'orica de la Materia Condensada, Universidad Aut\'onoma de Madrid, E-28049, Spain.}

\author{E.~del Valle}
\affiliation{Departamento de F\'isica Te\'orica de la Materia Condensada, Universidad Aut\'onoma de Madrid, E-28049, Spain.}

\author{F.~P.~Laussy}
\affiliation{Faculty of Science Engineering, University of Wolverhampton, Wulfruna St, WV1~1LY, United Kingdom.}
\affiliation{Russian Quantum Center, Novaya 100, 143025 Skolkovo, Moscow Region, Russia.}
\email{F.Laussy@wlv.ac.uk}

\begin{abstract}
  We map the Hilbert space of the quantum Harmonic oscillator to the
  space of Glauber's $n$th-order intensity correlators, in effect
  showing ``the correlations between the correlators'' for a random
  sampling of the quantum states. In particular, we show how the
  popular $g^{(2)}$ function is correlated to the mean population and
  how a recurrent criterion to identify single-particle states or
  emitters, namely $g^{(2)}<1/2$, actually identifies states with at
  most two particles on average. Our charting of the Hilbert space
  allows us to capture its structure in a simpler and physically more
  intuitive way that can be used to classify quantum sources by
  surveying which territory they can access.
\end{abstract}
\date{\today} \maketitle

\section{Introduction}

The formalization of quantum mechanics in the early days of its
construction~\cite{neumann_book32a} led to the introduction of the
Hilbert space as the structure to accommodate and unify the rules of
Heisenberg to compute observables~\cite{heisenberg25a} and the
wavefunction of Schr\"odinger to describe the quantum
states~\cite{schrodinger26b}. To Einstein's reported observation that
it would be enough to understand the electron, Dirac replied that ``it
would be enough if students could understand the harmonic
oscillator''~\cite{bloch_book97a}. This object indeed ranks as the
foundation for much of our description of the world, not so much as
the (quantum) mechanical object itself, but as the single mode of a
bosonic field. In this way, light can be described as a collection of
coupled harmonic oscillators, and such a basic notion as ``coherence''
was revolutionized in this workframe, changing from ``a monochromatic
field'' (a single oscillator is excited) to ``uncorrelated photons''
(regardless of their origin)~\cite{glauber06a}. In a modern
understanding, a single mode of well defined frequency can be chaotic
and a broadband, even time-varying field, can be coherent. It is only
because of the observed correlation in the physical observables
between thermal and/or chaotic fields with broad linewidths that the
identification of the two concepts came to be, that is still enduring
to this day. With technology and the rising of photonics, however, the
family of quantum states of the light field has been enlarged
considerably with more examples to distinguish these two concepts than
to associate them.  In quantum optical terms, coherence is nowadays
described by the Glauber correlators (we shall consider henceforth a
single bosonic mode~$a$ only):
\begin{equation}
  \label{eq:miemar23095934CET2016}
  g^{(n)}\equiv\langle a^{\dagger n}a^n\rangle/\langle\ud{a}a\rangle^n\,,
\end{equation}
where~$a$ is the annihilation operator (or ``ladder operator'') that
removes one quantum from the Fock state~$\ket{n}$ according to
$a\ket{n}=\sqrt{n}\ket{n-1}$.  The Glauber correlators are normalized
quantities obtained from ratio of observables:
\begin{equation}
  \label{eq:lunjun27110030CEST2016}
  G^{(n)}\equiv\langle{a}^{\dagger n}a^n\rangle\,.
\end{equation}
As we will use the normalized form in our discussion, instead of
$g^{(1)}$ which is unity, we will use as first-order variable the
normalization itself, that is the mean population of the oscillator
(average number of quanta):
\begin{equation}
  \label{eq:miemar23100404CET2016}
  n_0\equiv G^{(1)}=\langle\ud{a}a\rangle\,.
\end{equation}
These observables provide an essentially comprehensive description of
the quantum state of an harmonic oscillator, through its $n$-particle
fluctuation properties. In this text, we consider the states only and
not their dynamics according to some Hamiltonian and/or Liouvillian
equation of motion, so that all the correlators are same-time
correlators. In a dynamical context,
$g^{(1)}(\tau)\equiv\langle\ud{a}(0)a(\tau)\rangle/n_0$, becomes an
important observable by itself (its decay time from unity is related
to spectral coherence, that is, its departure from a single line).
The $g^{(n)}$ correlators describe collective fluctuations at several
orders, for instance, $g^{(2)}$ (the most widely used one) is related
to the variance of the population according to
$g^{(2)}=1+(\mathrm{Var}(n_0)-n_0)/n_0^2$. For Poisson fluctuations of
the population, $\mathrm{Var}(n_0)=n_0$ and~$g^{(2)}=1$. The
underlying quantum state is the coherent state~\cite{zhang90a}
theorized by Sudarshan~\cite{sudarshan63a} and
Glauber~\cite{glauber63a}.  Sub-Poissonian fluctuations are
characteristic of genuine quantum states of the field, i.e., with no
classical analogues, epitomized by the Fock state~\cite{lvovsky01a}.
Chaotic light, on the contrary, exhibits large fluctuations,
with~$g^{(2)}=2$. The underlying quantum state is the thermal density
matrix~\cite{glauber63a}. These correlators are also popularly known
as the ``$n$th-order quantum coherence functions''.

\section{Charting the Hilbert space}

All our discussion so far has been well-known introductory material to
quantum mechanics courses. In the following, we will study quantum
states of the harmonic oscillator (that can be thought of as the
single mode of a cavity) that go beyond the well known particular
cases through which we usually perceive the Hilbert space. The
canonical basis for the space is provided by the Fock
states~$\ket{n}$. While we will ultimately be concerned with the
complete space~$\mathcal{H}_\infty$ of the Harmonic oscillator, it
will be convenient to approach it through subpsaces of at most~$N$
quanta:
\begin{equation}
  \label{eq:marfeb16170000CET2016}
  \mathcal{H}_N=\left\{\sum_{k=0}^{N} \alpha_k\ket{k}~;~\Big(\alpha_k\in\mathbb{C}\Big)\wedge\Big(\sum_{k=0}^N|\alpha_k|^2=1\Big)\right\}\,.
\end{equation}
It is well-known, since Pegg and Barnett's attempts to define a phase
operator~\cite{pegg88a}, that working in a truncated Hilbert space of
arbitrary high maximum particle-number~$N$ allows us to get access to
physical properties that become pathological in the
infinite-dimensional space. We likewise consider truncated spaces that
can later be enlarged in a limiting process, in which
case~$\mathcal{H}_\infty\equiv\bigcup_{N=0}^\infty\mathcal{H}_N$.

While Eq.~(\ref{eq:marfeb16170000CET2016}) provides a comprehensive
depiction of $\mathcal{H}_N$, it is a deceiving picture that keeps
hidden much of the structure of the space. This is this structure
which we shall attempt to clarify in the following through its
visualization in terms of $g^{(n)}$ observables. The need for such an
analysis is motivated by the recent interest in exciting optical
targets with the new sources of quantum
light~\cite{lopezcarreno15a,lopezcarreno16a} made available by the
progress in quantum sources engineering~\cite{sanchezmunoz14a}. When
driving an harmonic oscilator with quantum light, one can bring the
system to a state that falls outside the known particular cases, even
though a considerable zoology has already been established. Indeed,
beyond the most famous cases already presented (thermal and coherent),
the literature describes a large family of quantum states for the
harmonic oscillators, with Gaussian states~\cite{wang07a},
predominantly squeezed states~\cite{breitenbach97a}, but also more
exotic families, such as cat states, i.e., superposition of coherent
states~\cite{leibfried05a} in various possible
combinations~\cite{dodonov74a}, two-photon coherent
states~\cite{yuen76a}, Fock-added coherent states~\cite{agarwal91b},
excited two-photon coherent states~\cite{xin96a} and their
generalization~\cite{wu04a}, binomial~\cite{stoler85a} and negative
binomial states~\cite{matsuo90a}, etc. The quantum world being such a
bizarre place, even such a simple operation as subtracting a state to
itself has inspired profuse discussions~\cite{othman16a}. In most
cases, the classifications follow from a particular scheme that allows
one to engineer the corresponding states.  As such, they do not
provide a picture of the Hilbert space that is both simple and
comprehensive and that would be practical to survey which regions of
the Hilbert space have already been covered, are the most easy of
access, which are its boundaries, if any, and what areas remain to be
explored.  This is such a picture that we provide based on the
particles joint-correlation properties.

A first simplification following from our approach that relies on
observables---Eqs.~(\ref{eq:miemar23095934CET2016}--\ref{eq:miemar23100404CET2016})---that
are sensitive to diagonal elements~$P_k\equiv|\alpha_k|^2$ only, is to
lift the distinction between \emph{pure states}, i.e., those of the
form of Eq.~(\ref{eq:marfeb16170000CET2016}) that can be written with
a wavefunction, and \emph{mixed states}, i.e., statistical
superpositions of these that are consequently of the type:
\begin{equation}
  \label{eq:psi}  
  \rho=\sum_{k=0}^{N} P_k\ket{k}\bra{k}+
  \sum_{k,l=0\atop k\neq l}^{N} P_{k,l}\ket{k}\bra{l}
  \,,
\end{equation}
with $P_{k,l}\in\mathbb{C}$ in general but $P_k\in\mathbb{R}$ (note
that we write~$P_k$ instead of~$P_{k,k}$). The mixed case is a
generalization which reduces to the pure one when
$P_{k,l}=\alpha_k\alpha_l^*$, and the second sum in Eq.~(\ref{eq:psi})
is redundant. A maximally mixed state on the other hand cancels
altogether the second sum. The arbitrary case interpolates between
these two situations corresponding to the degree of purity or
coherence (depending on terminology). We will leave it to context or
to cases of greater generality to decide which case is meant or
useful. For instance, $\alpha_n=\exp(-|\alpha|^2)\alpha^n/\sqrt{n!}$
can be understood as both the coherent state or as the random-phase
coherent state~\cite{glauber63c} (with all off-diagonal elements
zero). We will likewise use the notation
$\alpha_n=\sqrt{(1-\theta)\theta^n}$ (for~$0\le\theta\le1$) for both
the thermal state, which has null off-diagonal elements, or the pure
state version that is actually also of interest, as the eigenstate of
the Susskind-Glogower phase
operator~$(a\ud{a})^{-1/2}a$~\cite{susskind64a}, in which case it is
known as the ``coherent phase state''~\cite{shapiro91a} (for its
analogies with the coherent state, eigenstate of~$a$,
cf.~Ref.~\cite{hall93a} for a nice review).  In any case, the
important information for our exploration of the Hilbert space through
particle fluctuations resides in the first sum of Eq.~(\ref{eq:psi}).
The second sum can be summarized through a single number: the
coherence of the state.

In $\mathcal{H}_N$ where the total number of excitations is truncated,
$P_{N+m}= 0$ for $m \geq 1$ in Eq.~(\ref{eq:psi}), therefore, computing
the correlators~(\ref{eq:lunjun27110030CEST2016}) on the
states~(\ref{eq:psi}) yield the sequence:
\begin{subequations}
  \label{eq:lunjun27110304CEST2016}
  \begin{align}
    1 &= \sum_{n=0}^{N}  P_n\,, \\
    n_0 &=  \sum_{n=0}^{N} n P_n\,, \label{eq:miejul13193807CEST2016}\\
    G^{(2)} &= \sum_{n=0}^{N} n(n-1)P_n\,, \\
    &\vdots\nonumber \\ 
    G^{(N)} &= \sum_{n=0}^{N} n(n-1)\hdots (n-N+1) P_n\,.
  \end{align}
\end{subequations}

In this case, there is a bijection~$\mathcal{M}$ between the allowed
$G^{(n)}$ correlators and the states uniquely defined through the first
sum in Eq.~(\ref{eq:psi}). This can be written in matrix form:
\begin{equation}
  \label{eq:M}
  \vec{G}= \mathbb{M} \vec{P}
\end{equation}
between the vectors of $(N+1)$ elements $\vec{P}=(P_0,\cdots,P_N)^T$ and
$\vec{G}= ( 1, n_0, G^{(2)},...,G^{(N)} )^T$ with:
\begin{equation}
  \label{eq:MatrixForm}
  \mathbb{M}=
  \begin{pmatrix}
1 & 1 & 1 & \hdots & 1 & 1 \\
0 & 1 & 2 & \hdots & N-1 & N \\
0 & 0 & 2 & \hdots & (N-2)(N-1)   & N(N-1) \\
\hdotsfor{6} \\
0 & 0 & 0 & \hdots & 0 & N(N-1) \cdots 1
\end{pmatrix}\,,
\end{equation}
%
%
which, being upper-triangular, allows us to solve Eq.~(\ref{eq:M}) by
backward Gaussian substitution:
\begin{multline} 
  P_{N-k}=\\
  \frac{1}{(N-k)_{N-k}} \left\lbrace G^{(N-k)} - \sum_{k'=N-k+1}^{N} (N-k)_{k'} P_{k'} \right\rbrace \,,
\end{multline}
where
$(n)_k = \prod_{p=0}^{k-1} (n-p)= \expval{a^{\dagger\,k}
  a^k}{n}$.
This result can be expressed in an explicit recursive form by
developing all the coefficients:
\begin{equation}
  \label{eq:Param}
    P_i= \sum_{j \geq i}^N (-1)^{i+j} \frac{G^{(j)}}{i! (j-i)!}\,,
\end{equation}
for $0\le i\le N$. This is the inverse relation of
Eq.~(\ref{eq:lunjun27110304CEST2016}).

The expression holds true for any~$N$, and in a limiting process gets
extended to the case $N \rightarrow \infty$.  This relation can also
be obtained through the method of generating
functions~\cite{barnett_book97}. Now that this relationship
between~$P_k$ probabilities and~$G^{(n)}$ correlators is settled, we
are ready to approach the Hilbert space through the $g^{(n)}$
observables. Namely, we consider how a distribution of states
from~$\mathcal{H}_N$ is mapped in the space charted by~$g^{(n)}$. We
will call the latter space~$\mathcal{G}_N$.  Given that
$0 \leq P_n \leq 1$ for all~$n$, and their sum being unity by
normalization, one can foresee constraints for the correlators, if
only at the level of ``correlations between the correlators'', e.g.,
are they all large or small together? Or is it on the opposite
possible to have arbitrary high values of~$g^{(3)}$ for vanishing
$g^{(2)}$? And if so, are such states in ``equal numbers'' than those
of the opposite trend?  We answer these questions by providing the
density of states in the correlator space~$\mathcal{G}_N$. Namely, we
want to know how a distribution of points in~$\mathcal{H}_N$ is mapped
into~$\mathcal{G}_N$.

\begin{figure}[t]
  \centering
  \includegraphics[width=.6\linewidth]{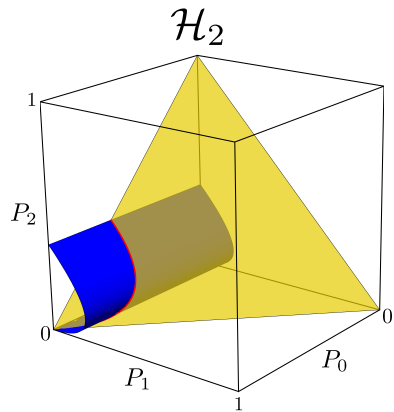}
  \caption{The two-particle Hilbert space $\mathcal{H}_2$ in the
    canonical Fock basis $P_k$ of probabilites for the state~$\ket{k}$
    is mapped on an equilateral triangle (yellow). A uniform sampling
    in this triangular space is a good representation of all the
    possible quantum states with at most two particles. The blue
    surface shows the states of constant $g^{(2)}$
    (namely~$g^{(2)}=1.3$) so that its intersection with the yellow
    triangle, shown as the red line, captures all the normalized
    physical quantum states with the corresponding~$g^{(2)}$. The
    ``measure'' (here length) of this line correspond to their density
    in the space.}
  \label{fig:h2plane}
\end{figure}

Since the Fock states basis Eq.~(\ref{eq:marfeb16170000CET2016}) is
intuitive, it is natural to consider a uniform distribution
in~$\mathcal{H}_N$ as a fair representation of all the quantum
states. For instance, the Hilbert space~$\mathcal{H}_2$ is a
2D~triangle in the~3D space $(P_0,P_1,P_2)$ (see
Fig.~\ref{fig:h2plane}) and all the quantum states of at most two
particles can be conveniently represented by the uniform distribution
over this geometry, namely, a constant distribution of value
$2/\sqrt{3}$ (the inverse area of an equilateral triangle of
side~$\sqrt{2}$). If a point is sampled randomly from this space,
corresponding to choosing one of the quantum states of the
form~$\alpha_0\ket{0}+\alpha_1\ket{1}+\alpha_2\ket{2}$ with the same
probability as any other, we then ask what is the probability that
this state will have a given population and second-order correlation
(all higher orders are zero since such states have at most two
particles). From Eqs.~(\ref{eq:lunjun27110304CEST2016}), it is easy to
see that the population lies between 0 and~2, and also that
$0\le G^{(2)}\le 2$, both maximised when~$P_2=1$ and all
other~$P_n=0$. It is not difficult, though less immediate, to show
that~$g^{(2)}$ is positive but unbounded (the possibility for two
particles to exhibit arbitrary large superbunching is also known from
the dynamics of bosonic cascades~\cite{liew16a}). Mathematically, this
means that $n_0$ and $G^{(2)}$ can vary independently between 0
and~2. To know if there is some degree of correlation between them, we
consider the distribution of states in the $(n_0, g^{(2)})$ space.

The quantum states with a given~$n_0$ are found as the intersection
between the triangle of normalized states in $(P_0,P_1,P_2)$ with the
plane of equation~$P_1+2P_2-n_0=0$. Similarly, the states with a given
$g^{(2)}$ are the intersection of the same supporting triangle with
the ellipsoid~$(P_1+2P_2)^2-2P_2/g^{(2)}=0$, shown as the blue surface
in Fig.~\ref{fig:h2plane}. The constant~$g^{(2)}$ states
in~$\mathcal{H}_2$ are consequently those identified by the red line
in Fig.~\ref{fig:h2plane}. This turns the question of the density of
states in~$\mathcal{G}_N$ into a problem of measuring surfaces in
hyperspaces: the measure, i.e., total area or volume related to a
certain manifold, has the same value regardless of which
parametrization (or metric) is chosen. This is tackled in differential
geometry with the first fundamental form $\mathbb{F}$, that provides
the trajectory in one space that is parametrically defined in the
other. The regions that are thus connected are, in general,
hypersurfaces. The relation reads:
\begin{equation}
  \label{eq:Fkk}
  \mathbb{F}_{k,k'}= \partial_{G^{(k)}} \vec{P}\cdot \partial_{G^{(k')}} \vec{P}
\end{equation}
where~$1\le k,k' \le N$ and~$\cdot$ is the scalar product between the
$\partial\vec P$ vectors.  As the transformation Eq.~(\ref{eq:M}) is
linear, the elements of $\mathbb{F}$ are constant, namely, they are
given by
$\mathbb{F}_{k,k'}=\sum_{i \leq k,k'} (-1)^{k+k'}\big/\big[i!^2 (k-i)!
(k'-i)!\big]$.
An element of (hyper)surface in $\mathcal{H}_N$ is related to the
corresponding element in $\mathcal{G}_N$ by
$\mathcal{P}_G dn_0\cdots dG^{(N)}= (\sqrt{| \mathbb{F}|}/A_N)
dP_0\cdots dP_N$
with~$\mathcal{P}_G$ the density of probability, $A_N$ is the volume
of the Hilbert space $\mathcal{H}_N$, that, being a simplex of
dimension~$N+1$, reads:
\begin{equation}
  \label{eq:SurfaceHn}
  A_N= \frac{\sqrt{N+1}}{N!}\, ,
\end{equation}
and the value of $|\mathbb{F}|$ can be computed from
Eq.~(\ref{eq:Fkk}) and is found in terms of the superfactorial
$\mathrm{sf}(N)= \prod_{i=0}^N i!$ as:
\begin{equation}
  \label{eq:fundamentalForm}
  \sqrt{|\mathbb{F}|}=\frac{\sqrt{N+1}}{\mathrm{sf}(N)}\, .
\end{equation}

While the computation is conveniently performed with~$\vec{G}$, we are
eventually interested in the space of normalized
correlators~$g^{(n)}$, that we will call~$\mathcal{g}_N$. A summary of
the spaces involved and the notations to identify them is given in
Table~\ref{tab:probabilities}.  There is another
bijection~$\mathcal{N}$ from $\mathcal{G}_N$ to $\mathcal{g}_N$, that
simply involves powers of~$n_0$ as
$( n_0,G^{(2)},...,G^{(N)} ) = ( n_0,n_0^2 g^{(2)},...,n_0^N g^{(N)}
)$.
The Jacobian for this transformation from $\mathcal{G}_N$ to
$\mathcal{g}_N$ reads:
\begin{align}
  \label{eq:J}
  J&=\left\vert \frac{\partial G^{(i)}}{\partial \g{j}} \right\vert =
  \begin{vmatrix}
    1 & 0 & 0 & 0 & \dots & 0 \\ 
    2n_0 g^{(2)} & n_0^2 & 0  & 0 & \dots & 0 \\
    3n_0^2 g^{(3)} & 0 & n_0^3 & 0 & \dots & 0 \\
    \hdotsfor{6} \\
    N n_0^{N-1}g^{(N)} & 0 & 0 & 0 & \dots & n_0^N
  \end{vmatrix}\notag
  \\
  &=\prod_{p=2}^N n_0^p
  =n_0^{(N^2+N-2)/2}\,.
\end{align}
This finally brings us to one of the main quantities of this text: the
joint density of
probability~$\mathcal{P}_g \left( n_0, g^{(2)},\cdots,g^{(N)}
\right)$.
Specifically, the probability that a state randomly picked
from~$\mathcal{H}_N$ has corresponding correlators~$n_0$, \dots,
$g^{(N)}$ in an infinitesimal hypervolume $dn_0\cdots dg^{(N)}$ is
$\mathcal{P}_g dn_0\cdots dg^{(N)}$.  Bringing all the results above
together, this density of probability is found as:
\begin{equation}
  \label{eq:miejun29164737CEST2016}
  \mathcal{P}_g \left( n_0, g^{(2)},\ldots,g^{(N)} \right) = \frac{n_0^{(N^2+N-2)/2}}{\mathrm{sf}(N-1)} \Theta({\mathcal{g}_N})\,,
\end{equation}
where~$\Theta(\mathcal{g}_N)\equiv\mathbf{1}_{\mathcal{N}\circ\mathcal{M}(\mathcal{H}_N)}$
is the support for the image of $\mathcal{H}_N$ through the
bijection~$\mathcal{N}\circ\mathcal{M}$, i.e., is~1 if there exists a
state with joint-correlators $n_0$, $g^{(2)}$, \dots, $g^{(n)}$ and
is~0 otherwise.  The
subset~$\mathcal{N}\circ\mathcal{M}(\mathcal{H}_N)$ remains to be made
explicit and its identification represents the core of the problem. It
is already notable, however, that, for physical states,
$\mathcal{P}_g$ is independent of all the correlators except the
population~$n_0$.

We now turn to particular cases to apply and illustrate these results.
In each case, the following procedure holds: a uniform distribution
of states in the space~$\mathcal{H}_N$ leads to a corresponding
distribution in~$\mathcal{g}_N$ given by
Eq.~(\ref{eq:miejun29164737CEST2016}).  The space~$\mathcal{g}_N$
itself is bounded when projected onto its~$n_0$ axis. The boundaries
are found from re-arranging the inequalities $0\le P_i\le 1$
with~$P_i$ given by Eqs.~(\ref{eq:Param}) to read as inequalities for
the correlators instead.  Marginal distributions can be obtained that
provide the distribution of quantum states in subspaces of interest
(e.g., $(n_0,g^{(n)})$).


\begin{table}[t]
  \begin{ruledtabular}
    \begin{tabular}{c|c}
      Subspace & Probability  \\
      \hline
      $\mathcal{H}_N$  & $\mathcal{P}(P_0,P_1,\ldots,P_N)=(1/A_N) \Theta\left(\mathcal{H}_N \right)$  \\
      $\mathcal{G}_N$ & $\mathcal{P}_G(n_0,G^{(2)},\ldots,G^{(N)})=(\sqrt{|\mathbb{F}|}/A_N) \Theta\left(\mathcal{G}_N \right)$ \\
      $\mathcal{g}_N$ & $\mathcal{P}_g(n_0,g^{(2)},\ldots,g^{(N)})=(J\sqrt{|\mathbb{F}|}/A_N) \Theta\left(\mathcal{g}_N \right)$  
    \end{tabular}
  \end{ruledtabular}
  \caption{
    Summary of notations for the various spaces
    introduced. $\mathcal{H}_N$ is the Hilbert space truncated to
    $N\in\mathbb{N}$ particles in the Fock basis. $\mathcal{G}_N$ is the
    corresponding space in the basis of unnormalized correlators~$G^{(N)}$
    and $\mathcal{g}_N$ in the space of Glauber correlators~$g^{(n)}$. The
    densities of probability are such that for a uniforma sampling $A_N$
    is the volume of the Hilbert space $\mathcal{H}_N$, $\mathbb{F}$ is
    the first fundamental form and $J$ the Jacobian of the transformation
    between the correlators and their normalized form. Their expressions
    are given in Eqs.~(\ref{eq:SurfaceHn}) and~(\ref{eq:fundamentalForm}).
    $\Theta$ is nonzero only if there is a physical state that provides
    the joint variables of the~$\mathcal{P}$ functions.
  }
  \label{tab:probabilities}
\end{table}

\section{The two-particle Hilbert space $\mathcal{H}_2$}

We consider first the simplest space distinct from that of the
two-level system ($\mathcal{H}_1$ is the Hilbert space of a qubit and
its complete characterization is textbook
material~\cite{nielsen_book00a}). Namely, $\mathcal{H}_2$, the space
spanned by $\ket{0}$ (vacuum), $\ket{1}$ and~$\ket{2}$, has
dimension~3 and can be fully represented geometrically in a 3D
Euclidean space. We have already used this space to illustrate the
nature of the Hilbert space in the~$P_k$ and $G^{(k)}$ bases in
Fig.~\ref{fig:h2plane}.

Equations~\eqref{eq:Param} read in this case:
\begin{subequations}
  \begin{align}
    \label{eq:miejul13212240CEST2016}
    P_0&=1- n_0 + \frac{n_0^2 g^{(2)}}{2}\,,\\
    P_1&=n_0(1 - n_0 g^{(2)})\,,\\
    P_2&=\frac{n_0^2 g^{(2)}}{2}\,,
  \end{align}
\end{subequations}
with $0 \leq P_k \leq 1$. 
%
%
\begin{figure}[t]
  \centering
  \includegraphics[width=\linewidth]{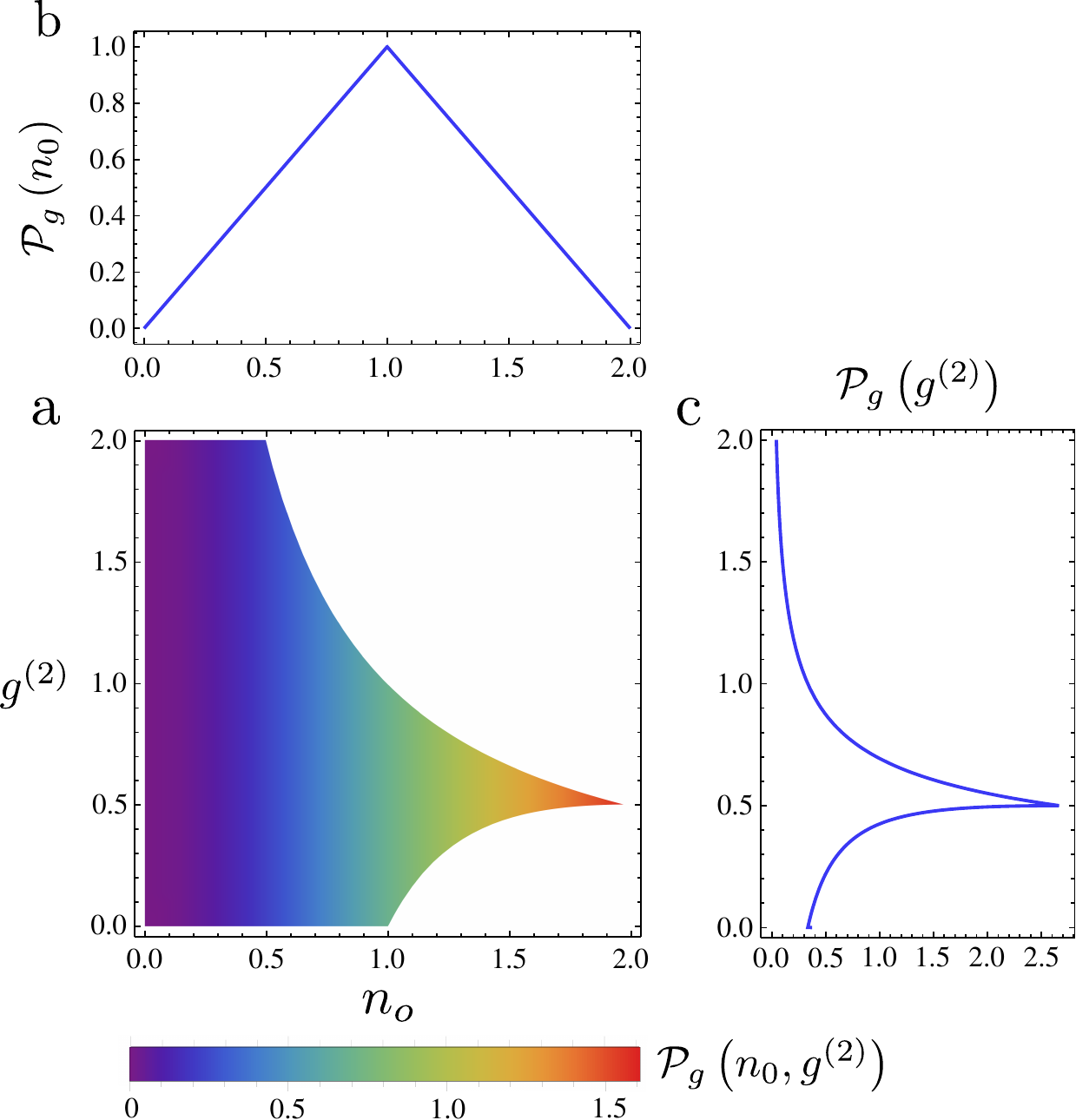}
  \caption{Charting of the Hilbert space $\mathcal{H}_2$ (up to
    two-particles). (a) Probability
    distribution~$\mathcal{P}_g (n_0,g^{(2)})$ of finding a quantum state
    with the corresponding population and~$g^{(2)}$ from a random
    sampling in~$\mathcal{H}_2$ (uniform distribution on the triangle
    in Fig.~\ref{fig:h2plane}). The space exhibits both a lower and
    upper boundary. (b) Distribution $\mathcal{P}_g(n_0)$ after
    averaging over~$g^{(2)}$ and (c) Distribution
    $\mathcal{P}_g(g^{(2)})$ after averaging over~$n_0$. The space is
    unbounded in~$g^{(2)}$, so that arbitrary high superbunching can
    be realized, what requires vanishing populations.}
  \label{fig:miemay4161007CEST2016}
\end{figure}
The reverse relations are familiar from the definitions of the
observables:
\begin{subequations}
  \begin{align}
    P_0+P_1+P_2 &=1\,, \\
    P_1+2 P_2 &= n_0\,, \\
    \frac{2 P_2}{(P_1 + 2P_2)^2} &= g^{(2)}\,.
  \end{align}
\end{subequations}

The corresponding joint probability~$\mathcal{P}_g$, i.e., the
probability of finding a state with given $(n_0, g^{(2)})$ from a
uniform sampling in the Hilbert space is:
\begin{equation}
  \label{eq:miejul13212432CEST2016}
  \mathcal{P}_g (n_0, g^{(2)}) = n_0^2 \ \Theta\left( \mathcal{g}_2 \right)\,,
\end{equation}
where $\Theta\left( \mathcal{g}_2 \right)$ vanishes if
$\left( n_0, g^{(2)} \right) \notin \mathcal{g}_2 $.  As already
stated, there is no explicit dependency of $\mathcal{P}_g$ on
$g^{(2)}$ once in $\mathcal{N}\circ\mathcal{M}(\mathcal{H}_2)$. Since
$\Theta(\mathcal{g}_2)$ is not everywhere one, there is, however, an
implicit dependency through $\mathcal{g}_2$'s geometry. This space is
found from the inequalities~(\ref{eq:miejul13212240CEST2016}) and can
be easily visualized as it is two-dimensional.  The inequalities
on~$P_k$ result in upper and lower boundaries for $n_0$ and~$g^{(2)}$:
\begin{subequations}
  \begin{align}
    g^{(2)}&\leq \frac{1}{n_0}\,, \label{eq:upperboundary}\\
    g^{(2)} &\geq \frac{\lfloor n_0 \rfloor (2n_0 - \lfloor n_0
      \rfloor -1)}{n_0^2}\,.\label{eq:lowerboundary}
  \end{align}
\end{subequations}
The lower boundary for $g^{(2)}$, Eq.~(\ref{eq:lowerboundary}) was
already known~\cite{lopezcarreno16a} and applies to
all~$\mathcal{H}_N$. There is also, however, an upper boundary,
Eq.~(\ref{eq:upperboundary}), that is specific to~$\mathcal{H}_2$.
Together, this constrains the states in~$\mathcal{H}_2$ to be confined
in the area shown in Fig.~\ref{fig:miemay4161007CEST2016}. The color
code there is that given by Eq.~(\ref{eq:miejul13212432CEST2016}), and
shows that states uniformly distributed in~$\mathcal{H}_2$ yield the
largest density of probability in the edge $(n_0,g^{(2)})=(2,1/2)$,
that is the point corresponding to~$\ket{2}$, since there is only one
state with this mean population and states with similar populations
also have a similar $g^{(2)}$. In contrast, there are many states with
mean population~1, but their range of $g^{(2)}$ is limited (between 0
and~1), the probability to find one of them is thus
intermediate. Finally, while there is also only one state with mean
population zero (the vacuum), states with similar populations can have
any positive~$g^{(2)}$, hence there is a small probability to find any
such state.  When disregarding the population, one finds that the
antibunching with highest probability is that of the Fock
state~$\ket{2}$, i.e., $g^{(2)}=1/2$, although another state will
likely have been drawn in its place. If it would be uncorrelated, it
would most likely have mean population~1.

The boundaries in $\mathcal{H}_2$ can also be written as:
\begin{equation}
  n_0 \leq \frac{1-\sqrt{1- 2g^{(2)}}\theta (1-2g^{(2)})}{g^{(2)}}
\end{equation}
where $\theta(x)$ is the Heaviside function. Regarding the upper
bound, for a given allowed population, $0\le n_0\le 2$, $g^{(2)}$
cannot be larger than $1/n_0$. The lesser the population, the greater
the maximum $g^{(2)}$ can be. This is consistent with results on
superbunching obtained from bosonic cascades~\cite{liew16a}, that show
that large bunching, $g^{(2)}\gg2$, develop as the system gets close
to vacuum.  Even though the joint probability takes a simple form, the
geometry of the Hilbert space when charted by the correlators has thus
a complex form. This echos in the reduced probability distributions,
that have a simple support, but inherit as a result complicated
functional expressions. Both distribution, $\mathcal{P}_g(n_0)$ and
$\mathcal{P}_g(g^{(2)})$, are obtained by integrating over the other
observable. The first one provides the population distribution:
\begin{equation}
\mathcal{P}_g(n_0) = \begin{cases}
n_0 & \text{if } 0 \leq n_0 \leq 1 \\
   2-n_0       & \text{if } 1 < n_0 \leq 2
\end{cases}\,,
\end{equation}
and the other one provides the $g^{(2)}$ distribution:
\begin{equation}
\mathcal{P}_g(g^{(2)}) = \begin{cases}
 \sqrt{\frac{8}{9}} \frac{(1-\sqrt{1-2 g^{(2)}}-g^{(2)})^{\frac{3}{2}}}{(g^{(2)})^3} & \text{if } 0 \leq g^{(2)} \leq \frac{1}{2} \\
   \frac{1}{3(g^{(2)})^3}      & \text{if } g^{(2)}> \frac{1}{2}
\end{cases}\,.
\end{equation}
Both distributions are piecewise functions, and are shown in
Fig.~\ref{fig:miemay4161007CEST2016}(b-c). A random sampling
in~$\mathcal{H}_2$ is thus most likely to produce a state with one
excitation if limiting to this observable, and an antibunching
of~$1/2$ if limiting to this observable. Jointly, however, the most
likely $g^{(2)}$ remains~1/2 but now for a population of~2. This does
not mean, however, that~$\ket{2}$ is most probably, only that states
close-by resemble it while states close-by, say, the vacuum, are very
different. 

The states that lie on the boundaries of the Hilbert space (we will
call them \emph{coin states}) are a superposition of two of the three
basis's Fock states:
\begin{equation}
  \label{eq:miejul13234351CEST2016}
  \sqrt{1-\frac{n_0}{2}} \ket{\mu} + \sqrt{\frac{n_0}{2}} \expo{i \theta} \ket{\nu}
\end{equation} 
with~$0\le\mu,\nu\le2$ such that~$\mu\neq\nu$. The
$\ket{0}$--$\ket{1}$ superpositions lie on the $x$-axis, the
$\ket{0}$--$\ket{2}$ define the upper boundary and the
$\ket{1}$--$\ket{2}$ define the lower boundary past~$n_0=1$. The
$n_0=0$ boundary is (set-topologically) open, that is, the states can
get asymptotically close to, but without touching, the boundary. Other
boundaries are closed since states~(\ref{eq:miejul13234351CEST2016})
are part of~$\mathcal{H}_2$. Note also that while $\mathcal{H}_2$ is
bounded, $\mathcal{g}_2$ is not, even though they are one-to-one
connected.

\section{The three-particle Hilbert space $\mathcal{H}_3$}

\begin{figure*}[thbp]
  \centering
  \includegraphics[width=.66\linewidth]{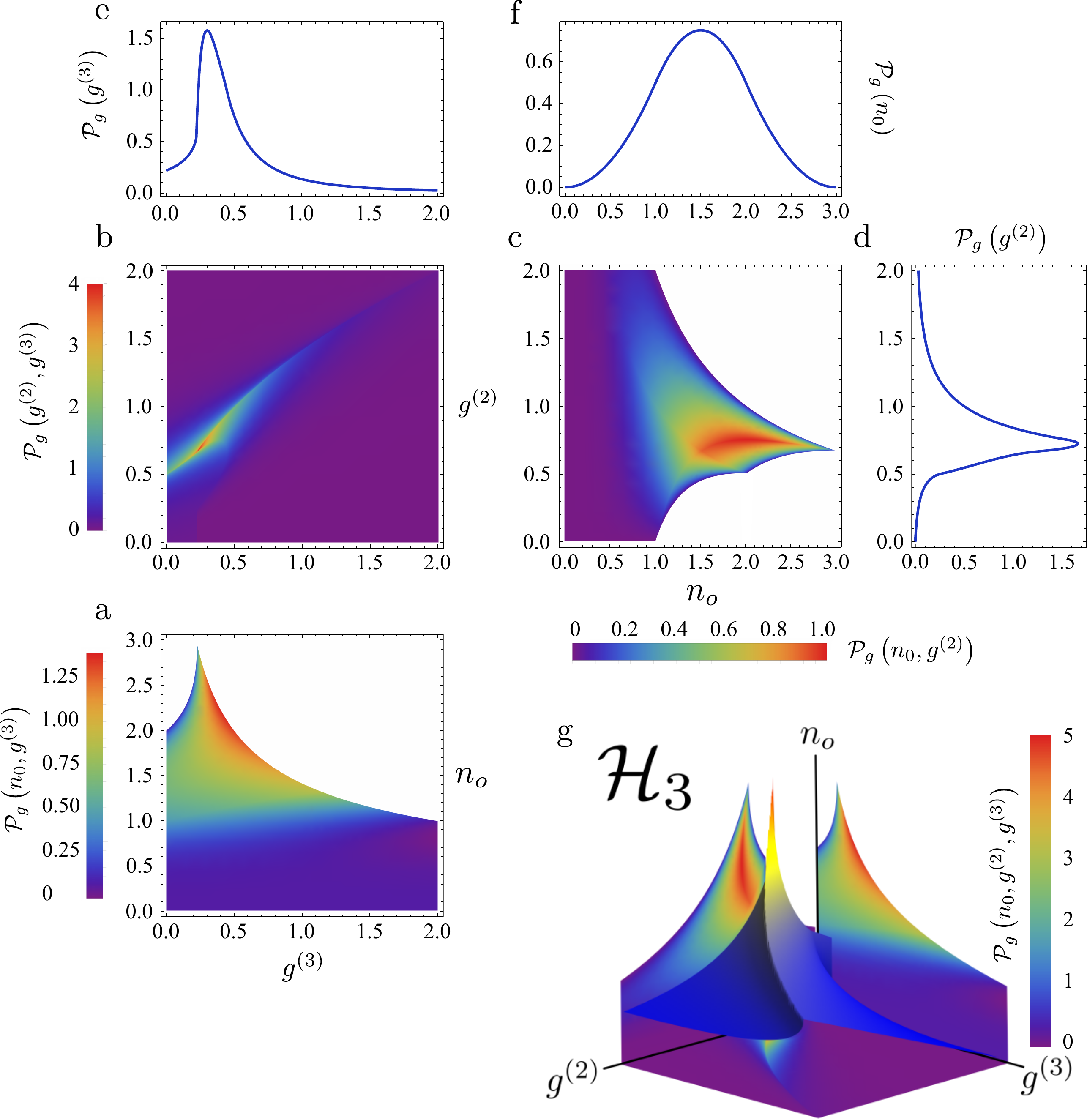}
  \caption{Charting of the Hilbert space $\mathcal{H}_3$ (up to
    three-particles). (a) Probability distribution~$\mathcal{P}_g
    (n_0,g^{(3)})$ after averaging over~$g^{(2)}$ of finding a quantum
    state with the corresponding population and~$g^{(3)}$ from a
    random sampling in~$\mathcal{H}_3$ (uniform distribution in an
    hypervolume [not shown]). The space exhibits both a lower and
    upper boundary similar to~$\mathcal{H}_2$ (note that it is
    rotated). (b) Distribution $\mathcal{P}_g (g^{(2)},g^{(3)})$ after
    averaging over~$n_0$. This subspace is unbounded unlike those that
    involve the population. (c) Distribution
    $\mathcal{P}_g(n_0,g^{(2)})$
    [cf.~Fig.~\ref{fig:miemay4161007CEST2016}] after averaging
    over~$g^{(3)}$. (d) Distribution $\mathcal{P}_g(g^{(2)})$, (e)
    $\mathcal{P}_g(g^{(3)})$ and (f) $\mathcal{P}_g(n_0)$. The latter
    distribution is of the Irwin-Hall type. (g) The complete
    distribution for~$\mathcal{H}_3$ lives in a 3D space, shown here
    through its upper boundary along with the three projections on 2D
    spaces.}
  \label{fig:miemay4160715CEST2016}
\end{figure*}

The principle for $\mathcal{H}_3$ is the same than for $\mathcal{H}_2$
but now in a 4D space, since the space is enlarged with a new
observable: the three-particle fluctuations $g^{(3)}$. This makes its
visualization trickier.  The results and their geometric
interpretation are still valid, but instead of 2D surfaces, one is now
dealing with hypersurfaces.

Equations~\eqref{eq:Param} read in this case:
\begin{subequations}
  \begin{align}
    P_0&=1- n_0 + \frac{n_0^2 g^{(2)}}{2} - \frac{n_0^3 g^{(3)}}{6}\,,\\
    P_1&=n_0 - n_0^2 g^{(2)}+ \frac{n_0^3 g^{(3)}}{2}\,,\\
    P_2&=\frac{n_0^2 g^{(2)}}{2} - \frac{n_0^3 g^{(3)}}{2}\,,\\
    P_3&= \frac{n_0^3 g^{(3)}}{6}\,,
  \end{align}
\end{subequations}
One can check that if $g^{(3)} = 0$ ($P_3 = 0$) then the structure of
$\mathcal{H}_2$ is recovered, as indeed $\mathcal{H}_2$ is a subspace
of $\mathcal{H}_3$. The distribution of states is found as:
\begin{equation}
  \mathcal{P}_g (n_0, g^{(2)}, g^{(3)}) = \frac{n_0^5}{2} \Theta \left(\mathcal{g}_3\right)\,,
\end{equation}
and as before, there is an explicit dependence only on the population,
with an implicit dependence on~$g^{(2)}$ and~$g^{(3)}$ from the fact
that the states are constrained to $\mathcal{g}_3$. The boundaries for
$(n_0 , g^{(2)} , g^{(3)})$ are complex. One can express them through
the constrains on one variable set by the two others. This yields,
for~$g^{(2)}$ as a function of~$n_0$ and~$g^{(3)}$:
\begin{subequations}
\label{eq:H3bound3}
\begin{align}
  g^{(2)}&\leq  \frac{n_0\g{3}}{2} + \frac{1}{n_0}\,,\label{eq:Mon29May105309BST2017} \\
  g^{(2)} &\geq \max\left( n_0 \g{3},\frac{n_0 \g{3}}{3}+\frac{2}{n_0}-\frac{2}{n_0^2} \right)\,,
\end{align}
\end{subequations}
and, for $g^{(3)}$ as a function of~$n_0$ and~$g^{(2)}$:
\begin{subequations}
\label{eq:H3bound2}
\begin{align}
 g^{(3)}&\le
   \min\left(\frac{g^{(2)}}{n_0},\frac{3 g^{(2)}}{n_0}-\frac{6}{n_0^2}+\frac{6}{n_0^3} \right)\,,\\
   g^{(3)}&\geq \max\left(0,\frac{2 g^{(2)}}{n_0}-\frac{2}{n_0^2}\right)\,,
\end{align}
\end{subequations}
with $0\le n_0\le3$ in both cases and $0\le g^{(2)}, g^{(3)}$ in
general. Observe from Eq.~(\ref{eq:Mon29May105309BST2017}) how the
bounding from~$1/n_0$ allows $g^{(2)}$ to grow arbitrarily for
vanishing populations. The equations apply for combinations of
$(n_0, g^{(3)})$ and $(n_0, g^{(2)})$ that are possible in the first
place, in which case the boundary for the third variable is as
indicated and consist of sharp inequalities, meaning that the equality
holds for some cases. If the combinations are not possible, the
equations as well may become impossible, requiring, e.g.,
$g^{(3)}<0$. The conditions for valid combinations define the
projected spaces $(n_0,g^{(N)})$ and will be given later
(cf.~Eqs.~(\ref{eq:juejul14001351CEST2016})) as they apply for
all~$N$. Note that in~$\mathcal{H}_2$ there is no such issue as the
projected space is also the full space.

The boundary set by~$g^{(2)}$ and~$g^{(3)}$ on~$n_0$ is the most
complicated one, although it is only bounding from above. It is given
in terms of two auxiliary functions, $f_1(g^{(2)},g^{(3})$ and
$f_2(g^{(3)})$, defined in the Appendix
(cf.~Eqs.~(\ref{eq:miejul13232528CEST2016}--\ref{eq:miejul13232539CEST2016})),
and reads:
\begin{equation}
  \label{eq:Tue23May175251BST2017}
  0\le n_0\leq U(g^{(2)}, g^{(3)})
\end{equation}
with
\begin{multline}
  \label{eq:UBoundN}
  U(g^{(2)}, g^{(3)})\equiv\\
  \begin{cases}
    \begin{array}{ll}
      f_1(g^{(2)},g^{(3)}) & \text{if} \ g^{(3)} \leq \frac{2}{9} \ \text{and} \ g^{(2)}<f_2 \left( \g{3} \right)\,, \\
      \frac{g^{(2)}-\sqrt{(g^{(2)})^2-2 g^{(3)}}}{g^{(3)}} & \text{if} \ g^{(3)} \leq \frac{2}{9} \ \text{and} \ g^{(2)}\geq f_2 \left( \g{3} \right)\,, \\
      \min[f_1, \frac{g^{(2)}}{g^{(3)}}] & \text{if} \ g^{(3)} \geq \frac{2}{9} \ \text{and} \ g^{(2)}<\sqrt{2 g^{(3)}}\,, \\
      \frac{g^{(2)}-\sqrt{(g^{(2)})^2-2 g^{(3)}}}{g^{(3)}} & \text{if} \  g^{(3)} \geq \frac{2}{9} \ \text{and} \ g^{(2)}\geq \sqrt{2 g^{(3)}}\,.
    \end{array}
  \end{cases}
\end{multline}
The value~$2/9$ comes from the $g^{(3)}$ of the Fock state~$\ket{3}$.

These results are already difficult to vizualize although still very
near the ground state of the oscillator. They are shown in
Fig.~\ref{fig:miemay4160715CEST2016} in the full~$\mathcal{H}_3$
space, where a single-sheet boundary encloses from above the space of
allowed states up to three particles. In most practical situations,
one is interested in pairwise correlations, so we address
$\mathcal{H}_3$ more systematically through its projections into its
subspaces.  This is obtained for any combination of two variables by
integrating over the third one.  The expressions are bulky and would
bring little enlightenment, so we keep them separate in the
Appendix. In this way, we can find $\mathcal{P}_g(n_0, g^{(2)})$
(Eq.~(\ref{eq:miejul13221300CEST2016})) and, for the new subspace now
accessible, $\mathcal{P}_g(n_0,\g{3})$
(Eq.~(\ref{eq:miejul13221243CEST2016})). The exact solutions have the
form of piecewise polynomial functions of their variables
(cf.~Appendix). It is interesting to compare
$\mathcal{P}_g(n_0, g^{(2)})$ for~$\mathcal{H}_3$ to that calculated
for~$\mathcal{H}_2$, where it was providing the complete picture,
while it is now averaged over~$g^{(3)}$. The boundaries are also
realized by Coin states of the form of
Eq.~(\ref{eq:miejul13234351CEST2016}), this time
with~$0\le\mu,\nu\le3$ (still with~$\mu\neq\nu$). This is true as well
for the new projected spaces $(n_0,g^{(3)})$.

As seen in Fig.~\ref{fig:miemay4160715CEST2016}, the Hilbert space is
bounded for the population but is not bounded when not involving this
parameter. This is due to intensity correlations of all orders being
largely independent from the population, thus allowing a normalizing
factor to make the quantity vanish or diverge (in contrast, $G^{(n)}$
are, like~$n_0$, all bounded).  As a result, all pairs of (positive)
values for $(g^{(2)}, g^{(3)})$ are possible. One can get antibunched
states of two particles that exhibit super three-particle bunching,
and reciprocally superbunching at the two-particle level but
three-particle antibunching, as well as, more expectedly, joint
two/three antibunching and superbunching, respectively. Making more
precise statements require to be more specific on how the correlators
reach their limits although one can be quite general regarding
vacuum. Table~\ref{tab:Tue23May175141BST2017} shows the upper
bound~$U$ for the population,
cf.~Eqs.~(\ref{eq:Tue23May175251BST2017}) and~(\ref{eq:UBoundN}), in
all the possible combinations for the limiting cases of $g^{(2)}$ and
$g^{(3)}$. If one correlator at least diverges, then the boundary
tends to~0, meaning that the state is dominated by vacuum,
$P_0\rightarrow1$. One can otherwise turn to the density of
probability for this subspace, that quantifies the relative occurence
of all possible combinations. It reads:
\begin{equation}
\mathcal{P}_g(g^{(2)}, g^{(3)})=\frac{1}{12} \left[U(g^{(2)}, g^{(3)}) \right]^6\,.
\end{equation}
and is shown in Fig.~\ref{fig:miemay4160715CEST2016}(b).

\begin{table}[!h]
\begin{center}
\begin{tabular}{c||c|c}
  $U(g^{(2)},g^{(3)})$ & $g^{(2)} \rightarrow 0$ & $g^{(2)} \rightarrow \infty$ \\
  \hline\hline
  $g^{(3)} \rightarrow0$, & {$\displaystyle 1 + \frac{g^{(2)}}{2} - \frac{g^{(3)}}{6}$} & {$\displaystyle\frac{1}{g^{(2)}} + \frac{2 g^{(3)}}{(g^{(2)})^3}$} \\
\hline
  $g^{(3)}\rightarrow\infty$ & {$\displaystyle\frac{g^{(2)}}{g^{(3)}}$} & 
$\begin{array} {lcl} i)&\displaystyle \frac{1}{g^{(2)}} + \frac{2g^{(3)}}{(g^{(2)})^3} \\ 
ii)&\displaystyle\frac{g^{(2)}}{g^{(3)}}
\end{array}$ \\
  \hline
\end{tabular}
\end{center}
\caption{Limiting cases of the population upper bound~$U(g^{(2)},g^{(3)})$ for all the possible combination of vanishing and diverging $g^{(2)}$ and~$g^{(3)}$. In the bottom right cell,  case i) applies to $(g^{(2)})^2>2g^{(3)}$ while case ii) applies to $(g^{(2)})^2<2g^{(3)}$. All cases except $g^{(2)}\rightarrow0$ and $g^{(3)}\rightarrow0$ lead to vanishing populations~$n_0$. Similar analyses could be undertaken for higher correlators.\label{tab:Tue23May175141BST2017}}
\end{table}

Integrating one step further, the probability distribution for~$n_0$
in~$\mathcal{H}_3$ can be obtained from either
equation~(\ref{eq:miejul13221300CEST2016})
or~(\ref{eq:miejul13221243CEST2016}) by integrating over the
extraneous variable, which yields:
\begin{equation}
  \mathcal{P}_g (n_0) = \begin{cases}
    \begin{array}{ll}
      \frac{n_0^2}{2} & \text{if} \ 0 \leq n_0 \leq 1\,, \\
      -\frac{1}{2} \left(2n_0^2 -6 n_0 +3 \right) & \text{if} \ 1<n_0 \leq 2\,, \\
      \frac{1}{2} \left(n_0^2 -6n_0 +9 \right) & \text{if} \ 2<n_0 \leq 3\,, \\
\end{array}
\end{cases}
\end{equation}
that is plotted in panel~f of Fig.~\ref{fig:miemay4160715CEST2016}.
In a similar way, one can obtain from
Eq.~(\ref{eq:miejul13221300CEST2016}) the reduced probability
distribution for~$g^{(2)}$ in $\mathcal{H}_3$, that is another bulky
expression (cf.~Eq.~\ref{eq:miejul13225823CEST2016}) expressed in
piecewise form with the distribution itself being, as in the other
cases, not only continuous but also everywhere differentiable.  We
could not find an analytical expression for~$\mathcal{P}_g(g^{(3)})$,
that is displayed in panel~e.

All the density of probabilities for all subspaces are shown in
Fig.~\ref{fig:miemay4160715CEST2016}. The density plots are also shown
as projections on their respective planes in the full 3D space. As one
can see, the structure of the Hilbert space is intricate.

\section{The $N$-particle Hilbert space $\mathcal{H}_N$}

Further analytical results are not convenient (we refer to the
Appendix as an illustration of how the exact solutions quickly become
cumbersome, already in~$\mathcal{H}_3$). From the three-particle
Hilbert space to higher dimensional ones, there is also a qualitative
step. The inequalities system can be handled for $\mathcal{H}_3$, in
which case, polynomials of degree~3 are involved and their roots admit
a closed form as given by Cardano--Tartaglia formula. For $N \geq 4$,
this method is not applicable (even if there exists the Ferrari
formula for fourth degree polynomials). Nevertheless, some general
characteristics can be inferred without closed-form solutions.

In all cases, the distributions for the population~$\mathcal{P}_g(n_0)$ follow
Irwin-Hall distributions (i.e., the distribution for the sum of $N$
independent random variables with a uniform distribution):
\begin{equation}
  \mathcal{P}_g (n_0) = \frac{1}{2(N-1)!} \sum_{k=0}^N (-1)^k  \binom {N} {k} (n_0 - k)^{n-1} \text{sgn}(n_0 -k)\,.
\end{equation}
As a result, for large~$N$, the distribution of population is normally
distributed. This result is actually trivial and follows directly from
our uniform sampling of the native Hilbert space (in the canonical
Fock basis).

One can also generalize to all~$N$, and thus also to the complete
harmonic oscillator Hilbert space, the boundaries of~$\mathcal{g}_N$,
that are constraining only when involving~$n_0$, in which case they
are given by (the proof is given in the Appendix):
\begin{subequations}
  \label{eq:juejul14001351CEST2016}
  \begin{align}
    g^{(k)} & \leq \frac{(N-1)!}{(N-k)!} \frac{1}{n_0^{k-1}}\,, \label{eq:juejul14000951CEST2016}\\
    g^{(k)} & \geq \frac{\lfloor n_0 \rfloor !}{(\lfloor n_0 \rfloor-k)!n_0^k} \left(1+ \frac{k(n_0-\lfloor n_0 \rfloor)}{\lfloor n_0 \rfloor+ 1 - k} \right)\,.
  \end{align}
\end{subequations}
These equations are the one that need being satisfied for~$N=3$ to
provide physical upper boundaries to
Eqs.~(\ref{eq:H3bound3}--\ref{eq:H3bound2}). As these are all sharp
inequalities, one can easily find in this way the maximum correlators
for a given population. 

\begin{figure}[thbp]
\centering
\includegraphics[width=\linewidth]{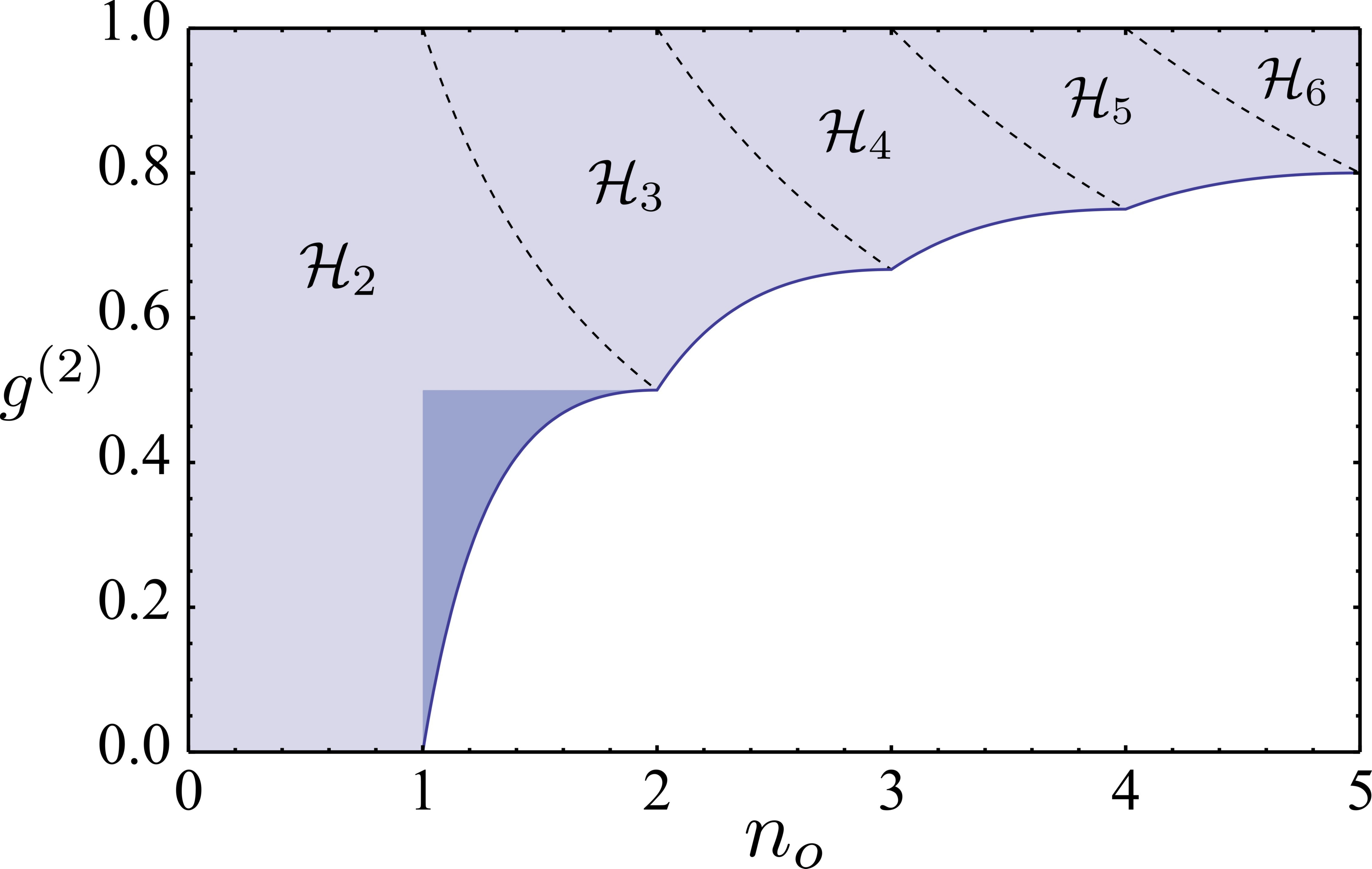}
\caption{Structure of the harmonic oscillator Hilbert space in the
  $(n_0,g^{(2)})$ subspace. There is a lower boundary but no upper
  boundary in~$\mathcal{H}_\infty$. The dashed lines show the upper
  boundaries that do exist for bunching in the truncated
  spaces~$\mathcal{H}_N$. The shaded region shows states with
  $\g{2}<0.5$ and $n_0>1$.}
\label{fig:NphaseSpace}
\end{figure}

Equations~(\ref{eq:juejul14001351CEST2016}) show that increasing~$N$,
the upper boundary for~$g^{(k)}$ wins territory in $\mathcal{g}_N$,
unlike the lower boundary. This has the effect of retaining only a
lower boundary in $\mathcal{H}_\infty$, that is shown for~$g^{(2)}$ in
Fig.~\ref{fig:NphaseSpace}. The dashed lines show the boundaries of
the successive~$\mathcal{H}_N$ spaces. The lower boundary of
Fig.~\ref{fig:NphaseSpace} shows that there exist states such
that~$n_0>1$ and $g^{(2)}<1/2$ (with
superpositions~$\sqrt{p} \ket{1}+\sqrt{(1-p)} e^{i \theta}\ket{2})$
for $0<p<1$ lying on the frontier). This is an important observation
as it invalidates a popular criterion in the literature that uses
$g^{(2)}<1/2$ as a criterion for single-particle states or, more
frequently, single-photon
emission~\cite{michler00a,dong07a,verma11a,dimartino12a,reimer12a,leifgen14a}.
Our map of the Hilbert space shows that the criterion~$g^{(2)}<1/2$ is
proper to identify states with less than two particles on average, not
one, as is the usual requirement for secure quantum protocols. The
actual criterion for the latter is $g^{(2)}=0$ and in the absence of
an exact mathematical zero, one should turn to other criteria for
single photon sources~\cite{arXiv_lopezcarreno16c}.

Note also that while any combination $\left(g^{(k)},g^{(k+n)} \right)$
(for $1\leq n \leq N-k$) is allowed, this imposes constrains on other
correlators, starting with~$n_0$ regardless of the truncation~$N$. In
fact, if $g^{(k)}=0$ for a particular $k$, it is easy to check that
every higher order correlator as well as every coefficient $P_n$ with
$n \geq k$ is also necessarily zero. This effectively truncates the
space. In the truncated space, not all combinations of correlators are
allowed even if they satisfy
Eqs.~(\ref{eq:juejul14001351CEST2016}). In~$\mathcal{H}_\infty$,
however, all combinations are allowed, with open boundaries of the
subspaces $\left( g^{(k)},g^{(k+n)} \right)$ when
$g^{(k)}, g^{(k+n)} \rightarrow 0$ (that is, excluding~0). A special
case is the limit $n_0 \rightarrow 0$ (that has already been mentioned
previously): no state except the vacuum, $\ket{0}$, has population
$n_0 = 0$.  This result agrees with the fact that the density of
probability $\mathcal{P}_N$ vanishes for $n_0=0$. A counterpart of
Table~\ref{tab:Tue23May175141BST2017} could be worked out. As the
details of how the population vanishes might not be of importance, we
only emphasize the following features that echo the results discussed
for~$\mathcal{H}_3$: in $\mathcal{H}_N$, if one correlator at least
diverges, then the state gets dominated by the vacuum:
$P_0\rightarrow1$. This can be seen from the fact that this
correlator, say of order~$k$, is bounded from above in the
$(n_0,g^{(k)})$ space, meaning that if $g^{(k)}\rightarrow\infty$ then
$n_0\rightarrow0$. If no correlator diverges, including the case of
all correlators vanishing, then the state can have a finite mean
population. This vanishing population for a diverging correlator is
not, however, true in general in~$\mathcal{H}_\infty$. There, one can
get diverging correlators for arbitrary large populations. Consider
for instance the case $(1-p)\ket{0}+p\ket{n}$, which, for any $M$ as
large as required, can be chosen to have population $np=M$ and
$g^{(2)}=n(n-1)p/(np)^2=(n-1)/(np)$ which tends to $1/p$ for $n$ large
enough. Thus, $g^{(2)}=1/p=n/M$ can be made as large as we want, by
considering~$n$ large enough (this is not possible if the space is
truncated). In turn, this makes $p$ very small, showing that in this
case again, the state is dominated by the vacuum, but the excited
state is now so largely populated that, on average, the population
does not have to vanish. There are other ways to arrive to similar
conclusions, showing in all that the structure of the Hilbert space is
a subtle one and that one should resist temptations of constraining
the quantum states from the behaviour of its Glauber correlators.

\begin{figure*}[thbp]
\centering
\includegraphics[width=\linewidth]{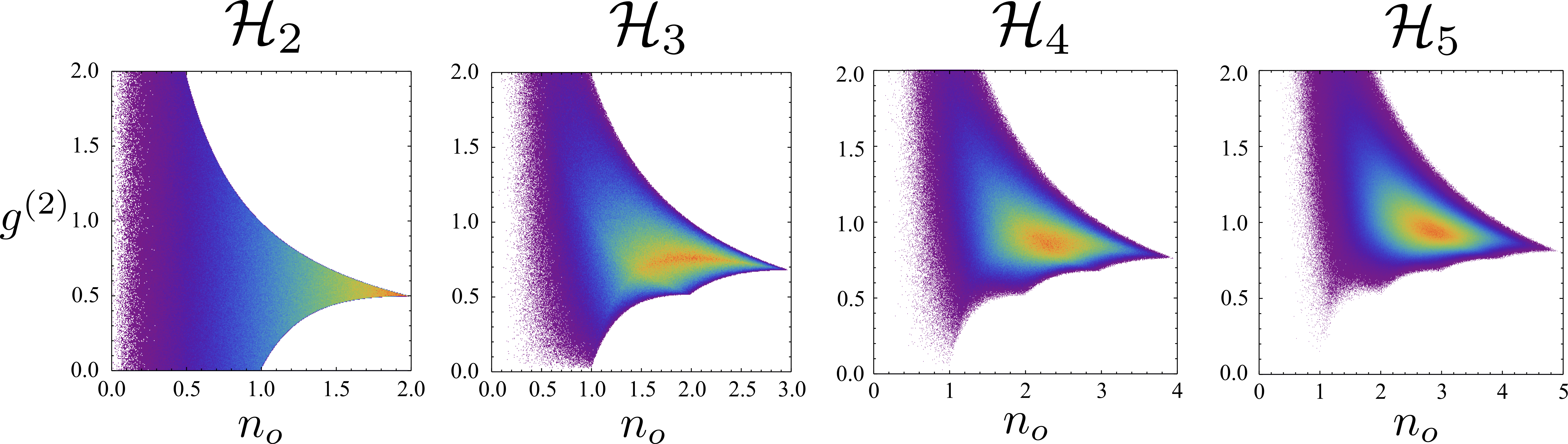}
\caption{Numerical distributions of states in the $(n_0,g^{(2)})$
  subspace as obtained from Monte Carlo uniform sampling in
  $\mathcal{H}_N$ for $2\le N\le 5$. The case~$N=2$ and~$3$ match with
  the analytical solutions presented above. Larger~$N$ weaken quantum
  features such as the scars observed clearly in~$\mathcal{H}_3$ that
  correspond to Coin states.}
\label{fig:juejul14003402CEST2016}
\end{figure*}

Finally, to gain an insight into higher-truncation spaces, we turn to
numerical methods. This also provides a way to check the analytical
results. In Fig.~\ref{fig:juejul14003402CEST2016}, we show the results
of Monte Carlo sampling of states in the Hilbert spaces
from~$\mathcal{H}_2$ till~$\mathcal{H}_5$ through their distribution
in the $(n_0,g^{(2)})$ subspace. The numerical results reconstruct
faithfully the distribution~$\mathcal{P}_g(n_0,g^{(2)})$ for the
case~$N=2$ and~$N=3$ for which we have provided analytical
solutions. It is also interesting that with increasing~$N$, one
observes a blurring of the quantum features such as the scars made by
the Coin states, clearly visible in~$\mathcal{H}_3$, faintly so
in~$\mathcal{H}_4$ and essentially gone in~$\mathcal{H}_5$, as well as
the kinky features of the the lower boundary. One witnesses in this
way the typical fading of quantum correlations with large number
of particles.

\section{Conclusions}

We have mapped the states of the Hilbert space of the harmonic
oscillator in the space of Glauber $n$th-order coherence
function~$g^{(n)}$, that capture the correlations of intensities at
various orders in the number of particles.  This allows to chart the
Hilbert space in a simple and visually appealing way, for instance in
the $(n_0,g^{(2)})$ subspace. We find that the Hilbert space has lower
boundaries when featuring the population, such that for populations
larger than one, some values of~$g^{(n)}$ become impossible (no
physical state can jointly provide them). There are no such
restriction when not involving the population, so that, e.g.,
arbitrary superbunching at some order can occur concurrently with
vanishing antibunching at some other order. For instance, one can find
states with $g^{(2)}\rightarrow 0$, $g^{(3)}\gg3!$,
$g^{(4)}\rightarrow 0$ and $g^{(5)}\gg5!$ (the values for
superbunching are defined with respect to the thermal fluctuations) or
indeed any combination, as long as the space has sufficiently high
truncation. In the truncated space, where~$N$ is finite, not all
combinations are possible as they need to belong to~$\mathcal{g}_N$
which has complicated upper boundaries. Said otherwise, given a
sequence of correlators that satisfy
Eqs.~(\ref{eq:juejul14001351CEST2016}), one can always produce
corresponding states as long as higher order correlators~$g^{(l)}$
for~$l>N$ can also be chosen (typically, nonzero). If they are forced
to be zero, they need to satisfy a more constrained condition
involving a $U$ function (cf.~\ref{eq:UBoundN} for the case~$N=3$). It
must be pointed out in particular that the region $g^{(2)} < 1/2$ and
$n_0>1$ is populated, which invalidates a popular criterion for
single-particle states or emission whenever~$g^{(2)}<1/2$. The
suitable such criterion is the simpler (and harder to achieve)
$g^{(2)}=0$.  What this criterion provides instead is a proof that the
emission has at most two particles on average. In summary, our results
provide a new, simple and practical representation of the possible
quantum states for the Harmonic oscillator, that should be of value
for instance to classify quantum sources by considering which areas of
the newly charted space they can reach.

\section{Acknowledgments}
We acknowledge funding from the European Union through the ERC
POLAFLOW and the Spanish MINECO under contract FIS2015-64951-R
(CLAQUE).

\bibliography{sahsbib}

\newpage

\appendix

\begin{widetext}
\section{Appendix A: Exact Results}

We list some of the exact and closed-form (but bulky) expressions for
quantities discussed or plotted in the main text. They are obtained
from the methods explained therein.

These are the auxiliary functions introduced to define the boundaries
for the population in the Hilbert space~$\mathcal{H}_3$  ($f_0$ is used
in~$f_1$):
\begin{equation}
\label{eq:miejul13232528CEST2016}
f_0 (\g{2},\g{3}) = \sqrt {6 (g^{(2)})^3 (g^{(3)})^2 - 
         3 (g^{(2)})^2 (g^{(3)})^2 - 18 g^{(2)} (g^{(3)})^3 + 
         9 (g^{(3)})^4 + 8 (g^{(3)})^3}\,,
\end{equation}

\begin{multline}
  f_1(g^{(2)},g^{(3)}) =-\frac {\sqrt[ 3] {-(g^{(2)})^3 + f_0
      (\g{2},\g{3}) + 3 g^{(2)} g^{(3)} - 3 (g^{(3)})^2}} {g^{(3)}} +{}\\
  \frac {18 g^{(3)} - 9 (g^{(2)})^2} {9 g^{(3)} \sqrt[ 3]
    {-(g^{(2)})^3 + f_0 (\g{2},\g{3}) + 3 g^{(2)} g^{(3)} 
      - 3 (g^{(3)})^2}} + \frac {g^{(2)}} {g^{(3)}}\,,
\end{multline}

\begin{multline}
\label{eq:miejul13232539CEST2016}
  f_2(g^{(3)})= \Re\left(\frac{\sqrt[3]{-8748 (g^{(3)})^2-4860
        g^{(3)}+8748 \left(g^{(3)}-\frac{2}{9}\right)^{3/2}
        \sqrt{g^{(3)}}+54}}{18 \sqrt[3]{2}} \right.
   \\
    {}\left.-\frac{-324 g^{(3)}-9}{9\times 2^{2/3} \sqrt[3]{-8748
        (g^{(3)})^2-4860 g^{(3)}+8748
        \left(g^{(3)}-\frac{2}{9}\right)^{3/2}
        \sqrt{g^{(3)}}+54}}\right)+\frac{1}{6}\,.
\end{multline}

These are reduced probability distribution in~$\mathcal{H}_3$:
\begin{equation}
  \label{eq:miejul13221300CEST2016}
  \mathcal{P}_g(n_0, g^{(2)}) = \begin{cases}
    \begin{array}{ll}
      3n_0^2-3n_0^3+\frac{3}{2}g^{(2)} n_0^4 & \text{if} \ g^{(2)}< \frac{3}{n_0}-\frac{3}{n_0^2} g^{(2)} \leq \frac{1}{n_0}\,, \\
      3n_0^2-2n_0^3+\frac{g^{(2)}n_0^4}{2} & \text{if} \ g^{(2)}< \frac{3}{n_0}-\frac{3}{n_0^2} 
      \  \text{and} \ g^{(2)} > \frac{1}{n_0}\,, \\
      \frac{g^{(2)} n_0^4}{2} & \text{if} \ g^{(2)} \geq \frac{3}{n_0}-\frac{3}{n_0^2} 
      \  \text{and} \ g^{(2)} < \frac{1}{n_0}\,, \\
      n_0^3-\frac{g^{(2)}n_0^4}{2} & \text{if} \ g^{(2)} \geq \frac{3}{n_0}-\frac{3}{n_0^2} 
      \  \text{and} \ g^{(2)} \geq \frac{1}{n_0}\,, \\
    \end{array}
  \end{cases}
\end{equation}

\begin{equation}
  \label{eq:miejul13221243CEST2016}
  \mathcal{P}_g(n_0,\g{3}) = 
  \frac{n_0^5}{2}
  \begin{cases}
    \begin{array}{ll}
      \frac{2}{n_0^2} - \frac{1}{n_0}+\frac{n_0 \g{3}}{6}  & \text{if} \ \g{3} < \frac{6-6n_0}{n_0^3 -3n_0}\,, \\
      -\frac{\g{3}}{n_0} + \frac{1}{n_0}+\frac{n_0 \g{3}}{2} & \text{if} \ \g{3} \geq \frac{6-6n_0}{n_0^3 -3n_0}\  \text{and} \ n_0 \geq 
\sqrt{3}\,, \\
      \frac{2}{n_0^2} - \frac{1}{n_0}+\frac{n_0 \g{3}}{6} & \text{if} \ \g{3} \geq \frac{6-6n_0}{n_0^3 -3n_0} 
      \  \text{and} \ n_0 < \sqrt{3}\,.
    \end{array}
  \end{cases}
\end{equation}

\begin{equation} 
  \label{eq:miejul13225823CEST2016}
  \begin{split}
    &\mathcal{P}_g (g^{(2)}) = \\
    & \begin{cases}
      \begin{array}{ll}
        \frac{2 g^{(2)} \left(\left(-12 \sqrt{9-12 g^{(2)}}+16 \sqrt{1-2 g^{(2)}}+75\right) g^{(2)}+63 \sqrt{9-12 g^{(2)}}-56 \sqrt{1-2 g^{(2)}}-190\right)-81 \sqrt{9-12 g^{(2)}}+48 \sqrt{1-2 g^{(2)}}+195}{60
        (g^{(2)})^4} & \text{if} \ 0 \leq g^{(2)} \leq \frac{1}{2}\,, \\
        \frac{2 g^{(2)} \left(-3 \left(4 \sqrt{9-12 g^{(2)}}-16 \sqrt{4-6 g^{(2)}}+75\right) g^{(2)}+63 \sqrt{9-12 g^{(2)}}-224 \sqrt{4-6 g^{(2)}}+370\right)-81 \sqrt{9-12 g^{(2)}}+256 \sqrt{4-6 g^{(2)}}-271}{60
        (g^{(2)})^4} & \text{if} \ \frac{1}{2}<g^{(2)} \leq \frac{2}{3}\,, \\
        \frac{2 \sqrt{9-12 g^{(2)}} (21-4 g^{(2)}) g^{(2)}-27 \sqrt{9-12 g^{(2)}}+5}{10 (g^{(2)})^4} & \text{if} \ \frac{2}{3}<g^{(2)} \leq \frac{3}{4}\,, \\
        \frac{1}{2 (g^{(2)})^4}  & \text{if} \ g^{(2)} > \frac{3}{4}\,.
      \end{array}
    \end{cases}
  \end{split}
\end{equation}
\end{widetext}

\section{Appendix B: Upper boundaries for $g^{(k)}$ in $\mathcal{H}_N$}

\begin{prop}
  Given some $\mathcal{H}_N$, for every pair $G^{(k-1)}$ and $G^{(k)}$
  with $k \leq N$, the inequality
  $(N-k+1)! G^{(k-1)} \geq (N-k)! G^{(k)}$ is satisfied. Subsequently,
  it holds that
  $ 0! G^{(N)} \leq 1! G^{(N-1)} \leq \cdots \leq (N-3)! G^{(3)} \leq
  (N-2)! G^{(2)}$.
\end{prop}

Since these observables can be expressed as:
\begin{subequations}
\begin{align}
G^{(k-1)} &= \sum_{n=k-1}^N \frac{n!}{(n-k+1)!} P_n \,, \\
G^{(k)} &= \sum_{n=k}^N \frac{n!}{(n-k)!} P_n \,,
\end{align}
\end{subequations}
it follows that:
\begin{equation}
  \begin{split}
    G^{(k-1)}-\frac{(N-k)!}{(N-k+1)!} G^{(k)} =(k-1)! P_{k-1} + {}\\
    \sum_{n=k}^N \left(\frac{1}{(n-k+1)!} - \frac{(N-k)!}{(N-k+1)! (n-k)!)}  \right)n! P_n\,.
\end{split}
\end{equation}

The term between parentheses in the summation is always greater than 0
and is equal to zero only if $n=N$. Therefore, the right side of the
last equation is greater than 0 as well:
\begin{equation}
  G^{(k-1)}-\frac{(N-k)!}{(N-k+1)!} G^{(k)} \geq 0\,,
\end{equation}
i.e., $(N-k+1)! G^{(k-1)} \geq (N-k)! G^{(k)}$.

\begin{prop}
  In every Hilbert space $\mathcal{H}_N$, $\g{2}$ admits an upper
  boundary, that is given by $\frac{N-1}{n_0}$.
\end{prop}

From the definition for $n_0=\sum_{n=0}^N n P_n$ and
$G^{(2)}= \sum_{n=0}^N n(n-1)P_n$ in $\mathcal{H}_N$, we find,
multiplying $n_0$ by $N-1$:
\begin{equation}
\label{eq:proof1}
\sum_{n=0}^N n(N-1) P_n=(N-1) P_1+\cdots+N(N-1)P_N\,.
\end{equation}
Substracting $G^{(2)}$ from expression \eqref{eq:proof1} leads to
$\sum_{n=0}^N n(N-n) P_n$.  This is always greater than 0 and only
equal if every term of the summation is null since all of them are
positive (remembering that $ 1\geq P_n \geq 0$). Therefore:
\begin{equation}
(N-1)n_0 \geq G^{(2)}\,,
\end{equation}
or, since $G^{(2)}=n_0^2 g^{(2)}$:
\begin{equation}
\label{eq:g2ineq}
g^{(2)} \leq \frac{N-1}{n_0}\,.
\end{equation}

Finally, $g^{(2)}$ can reach its upper boundary only if every $P_n$
vanishes excepting $P_0$ and $P_N$, i.e., when the corresponding state
is a ``Coin state'', cf.~Eq.~(\ref{eq:miejul13234351CEST2016}).  
Assuming both propositions, we can infer that:
\begin{equation}
G^{(k)} \leq \frac{(N-2)!}{(N-k)!} G^{(2)}\,.
\end{equation}
Furthermore, as $G^{(k)}$ can be written as $n_0^k \g{k}$ and from
Eq.~(\ref{eq:g2ineq}), we obtain:
\begin{equation}
\g{k} \leq \frac{(N-2)!}{(N-k)!} \frac{\g{2}}{n_0^{k-2}} \leq \frac{(N-1)!}{(N-k)!} \frac{1}{n_0^{k-1}}\,.
\end{equation}

\end{document}